\documentclass[aip,reprint]{revtex4-1}

\usepackage{graphicx}
\usepackage{ragged2e}
\usepackage{hyperref}
\usepackage{amsmath}
\usepackage[all]{nowidow}
\usepackage{soul}
\usepackage{xcolor}
\sethlcolor{yellow}
\usepackage[normalem]{ulem}

\draft 

\begin{document}


\title{Temporal characterization of femtosecond electron pulses inside
ultrafast scanning electron microscope} 



\author{Kamila Moriová}
 \email{kamila.moriova@matfyz.cuni.cz}
 \author{Petr Koutenský}
\author{Marius-Constantin Chirita-Mihaila}
\author{Martin Kozák}%
\affiliation{Department of Chemical Physics and Optics, Faculty of Mathematics and Physics,
Charles University, Ke Karlovu 3, 12116 Prague 2, Czech Republic
}%


\date{\today}

\begin{abstract}
In this work, we present the implementation of all-optical method for directly measuring electron pulse duration in an ultrafast scanning electron microscope. Our approach is based on the interaction of electrons with the ponderomotive potential of an optical standing wave and provides a precise \textit{in situ} technique to characterize femtosecond electron pulses at the interaction region across a wide range of electron energies (1–30~keV). By using single-photon photoemission of electrons by ultraviolet femtosecond laser pulses from a Schottky emitter we achieve electron pulse durations ranging from 0.5~ps at 30~keV to 2.7~ps at 5.5~keV under optimal conditions where Coulomb interactions are negligible. Additionally, we demonstrate that reducing the photon energy of the femtosecond pulses used for photoemission from 4.8~eV (257.5~nm) to 2.4~eV (515~nm) decreases the initial energy spread of emitted electrons, leading to significantly shorter pulse durations, particularly at lower electron energies.
\end{abstract}


\maketitle 

\section{Introduction}

In recent years, ultrafast electron microscopy (UEM) has emerged as a powerful method for investigating the dynamics of fundamental processes with high temporal and spatial resolution \cite{flannigan20124d}. Four-dimensional visualization of dynamic processes can be achieved through laser pump and electron probe experiments, where an optical pulse excites the sample, and an electron pulse probes the transient state of the material at precise time delay with respect to the arrival of the optical pump. Unlike traditional optical methods of pump-probe microscopy, which are spatially averaged over the probe spot size and constrained by the wavelength of light (Abbe's limit), UEM leverages the shorter wavelength of electron probes to offer unique insights at the atomic scale with significantly better spatial resolution. In the past, time-resolved imaging with pulsed electron beams has enabled studies of phenomena such as the dynamics of phase transitions \cite{baum20074d, gedik2007nonequilibrium, danz2021ultrafast, van2013single}, the kinetics of fast reactions in nanomaterials \cite{sinha2019nanosecond, ruan2007dynamics}, magnetization dynamics \cite{rubiano2018nanoscale, berruto2018laser} or the structural dynamics of molecules \cite{hensley2012imaging, ihee2001direct}. There is also growing interest in using this technique to study the optical response of nanophotonic structures \cite{polman2019electron, barwick2009photon, piazza2015simultaneous} or ultrafast charge dynamics at semiconductor interfaces for optoelectronic applications \cite{kosar2024time, cho2014visualization}. Additionally, recent advancements in UEM have expanded the focus from material studies to the investigation of electron-light interactions, enabling the manipulation of electron wavefunctions \cite{feist2015quantum, tsarev2023nonlinear, shiloh2022quantum, talebi2020strong}. These developments have led to applications such as the generation of attosecond electron pulses \cite{priebe2017attosecond, kealhofer2016all, kozak2018ponderomotive, ryabov2020attosecond, morimoto2018diffraction, morimoto2018attosecond}, electron acceleration \cite{chlouba2023coherent}, and the formation of electron vortex beams \cite{vanacore2019ultrafast, kozak2021electron}.

To achieve the highest time resolution in UEM, it is essential to generate short electron pulses and accurately determine their durations, as the temporal resolution of pump-probe experiments is given by the electron pulse duration rather than the detector's acquisition speed. Currently, two techniques are used to generate pulsed electron beams in UEM \cite{zhang2019photoemission}: laser-driven photoemission sources and electron beam blanking. These techniques can achieve electron pulse durations on the femtosecond and picosecond scales, respectively \cite{moerland2016time}. Furthermore, a design for a beam blanker incorporating a photoconductive switch has been proposed, which could enable electron pulse widths as short as 100~fs \cite{weppelman2018concept, weppelman2019pulse}. However, maintaining the short duration of electron pulses from source to sample is challenging due to the velocity spread of electrons and the action of Coulomb forces. Electron-electron repulsion by electrostatic forces can be mitigated by using less than one electron per pulse on average. Pulse broadening due to dispersive propagation of electrons with finite energy-momentum spread is unavoidable and leads to prolongation of the electron pulse duration to several hundreds of femtoseconds \cite{baum2013physics, aidelsburger2010single}. 

Only few techniques have been developed to directly measure the sub-picosecond durations of low-density electron pulses. One approach is based on electron-light cross-correlation utilizing the inelastic scattering of electrons induced by the interaction with optical near field excited at the surface of a nanostructure \cite{barwick2009photon, park2010photon, park2012chirped, feist2017ultrafast, plemmons2014characterization, kozak2018ultrafast, kirchner2014laser}. However, this technique is limited by the requirement for the UEM or electron gun to be equipped with a high-resolution spectrometer capable of spectrally filtering inelastically scattered electrons. Additionally, experiment must be designed for a specific electron energy, as the interaction between the electrons and the laser field is energy-dependent. Alternatively, the electron pulse can be characterized by streaking in a THz field generated by high-density photoemitted electrons from a transmission electron microscopy (TEM) grid \cite{plemmons2017ultrafast} or directly by the THz waveform \cite{kealhofer2016all} but this method has limited time resolution.

\begin{figure*}[t]\centering
\includegraphics[scale=0.34]{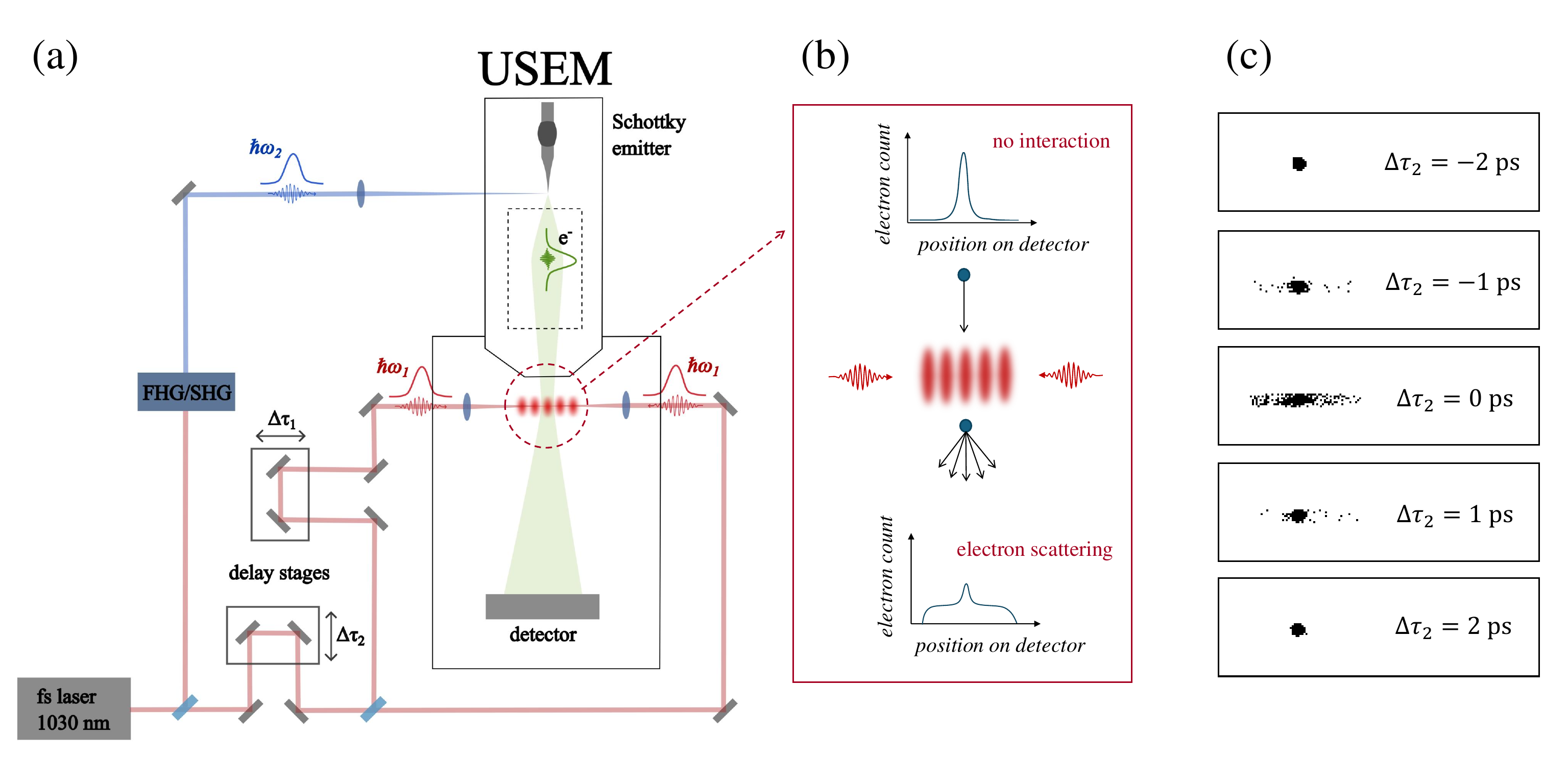}
\caption{\justifying
(a) Layout of the experimental setup. Electron pulses are photoemitted from a Schottky nanotip using focused ultraviolet or visible laser pulses (the fourth or second harmonic of a femtosecond laser with ytterbium-doped gain medium with a central wavelength of 1030~nm). The electron pulses interact with an optical grating formed by two overlapping laser beams, resulting in scattering by the ponderomotive force. Precise timing of the electron and laser pulses is achieved using two independent delay lines, enabling control over their relative arrival times. (b) The schematics depicts the process of elastic ponderomotive scattering, where electrons interact with a standing light wave in a classical interaction regime. (c) Images of scattered electron beam captured on the detector at various laser pulse–electron pulse delays, with optimal overlap defined as $\Delta \tau_2 = 0$~ps. The images were acquired for 15~keV electron pulses under conditions of negligible Coulomb repulsion, using a 64~$\mu$m aperture. The data represent an integration over 25,000 pulses, superposed across five measurements.
}
\label{fig1}
\end{figure*}

The techniques described above rely on high-resolution electron spectrometers that are currently optimized for high-energy electrons, as commonly used in ultrafast transmission electron microscopes (UTEMs). In contrast, scanning electron microscopes (SEMs) operate at much lower electron energies (usually 1–30 keV), which are designed for imaging of sample surface using secondary or back-scattered electrons. The low-energy electrons typically do not propagate through the sample in SEM and thus there is no transmitted electron beam, for which the inelastic interaction with light can be characterized. As a result, implementing electron energy loss spectroscopy (EELS)-like techniques in the low-energy electron regime is not yet widely adopted. Here, we present the integration of an alternative all-optical method that enables the direct temporal characterization of femtosecond electron pulses within an ultrafast scanning electron miscroscope (USEM), where determining electron pulse durations has mostly relied on theoretical predictions \cite{qian2002electron, gahlmann2008ultrashort}. This approach, based on elastic electron scattering mediated by the ponderomotive potential of an optical standing wave, offers an efficient and precise method for measuring electron pulse durations across a wide energy range, without requiring modifications to the experimental setup \cite{hebeisen2008grating, hebeisen2009thesis}. While similar techniques have been demonstrated in earlier works \cite{hebeisen2008grating, hebeisen2009thesis, gao2012full, chatelain2012ultrafast, tokita2015strong}, they have not yet been applied to characterize pulse durations over a broad energy range in a USEM. Implementing direct temporal characterization within a USEM allows for further optimization of experimental parameters, ultimately enabling the generation of shorter electron pulses.

\section{Methods}

\subsection{Experimental setup}

In our experimental setup, illustrated in FIG. \ref{fig1}(a), we generate electron pulses by laser-triggered emission from a Schottky emitter in Verios 5 UC SEM manufactured by ThermoFischer Scientific (energy range 1 to 30~keV) by illuminating the apex of the emitter tip with laser pulses with central wavelengths of 257.5~nm (photon energy of 4.8~eV) or 515~nm (2.4~eV). We aim to reach the optimal emission regime to achieve low-density electron pulses, eliminating space-charge effects and preventing Coulomb broadening during propagation while maintaining sufficient electron signal. This is achieved by reducing the photoemission laser power, which leads to a lower beam current while maintaining low initial energy and temporal spreads of the electrons\cite{yanagisawa2011energy}. Changing the aperture diameter does not produce any significant effect on the electron pulse duration, as the main temporal broadening occurs close to the emitter nanotip. The reason is that the Coulomb interaction scales as $ 1/{r^2}$ and the distance between electrons $r$ in a pulse increases with their average velocity in the laboratory frame. The electrons are thus the closest to each other during or right after photoemission, when their energy and velocity are still low. In our measurements we utilize laser repetition rates of 50~kHz and 500~kHz, with the latter providing an improved signal-to-noise ratio due to 10-times higher current.

Inside the SEM chamber we implement an optical setup for the generation of optical grating with two counter-propagating laser pulses with central wavelength of 1030~nm and duration of $\tau_l=220$~fs. The pulses are focused to a spot size of 10~$\mu$m ($1/e^2$) approximately 9~mm downstream the pole piece. Convergence angle of the beam was approximately 1~mrad. The temporal overlap of the two laser pulses and the electron pulse within the interaction region is achieved by controlling their relative arrival times using two independent optical delay lines. Spatial overlap is ensured by aligning both laser beams and the electron beam at the nanotip positioned at the center of the electron column, which is removed during the experiments. The scattered electron signal is detected by a hybrid-pixel detector Timepix3. Images of the scattered electron beam for different electron pulse-laser pulse delays are shown in FIG. \ref{fig1}(c). The duration of the electron pulse is reconstructed from the measured scattered electron signal across these time delays, as described below in section \ref{puldur}.

We note that for certain applications, the optical setup for measuring electron pulse duration—implemented inside the chamber—can remain in place, allowing \textit{in situ} characterization during SEM operation. In our experiment, however, the sample stage must be rotated to an angle of $60^\circ$ to enable electron transmission through both the optical standing wave and the sample area. When the sample is mounted in a specialized holder, it can be retracted to perform the pulse duration measurement and then reinserted for subsequent ultrafast measurements with electron pulses.

\subsection{Electron pulse duration characterization}
\label{puldur}

An approach allowing characterization of electron pulses is elastic ponderomotive scattering of a single high-intensity pulsed laser beam \cite{hebeisen2006femtosecond}. However, this method is unsuitable as an \textit{in situ} technique for UEM due to the high laser pulse energy requirements, which exceed the capabilities of conventional femtosecond lasers. An alternative is to enhance electron scattering using an optical standing wave created by two counter-propagating pulsed optical beams of the same frequency. The reason for a strong enhancement of the electron scattering probability is the fact that the classical ponderomotive force depends on the spatial gradient of optical fields, which is strongly enhanced in the optical standing wave due to the interference modulation of the light intensity. The concept of employing standing light waves as an optical grating for electrons was initially suggested by Kapitza and Dirac \cite{kapitza1933reflection} and was later experimentally demonstrated for the scattering \cite{bucksbaum1988high} and diffraction \cite{freimund2001observation} of low-energy electrons. Optical gratings are also employed in matter-wave interferometry for molecules and clusters \cite{haslinger2013universal}.

\begin{figure*}[t]\centering
\includegraphics[scale=0.34]{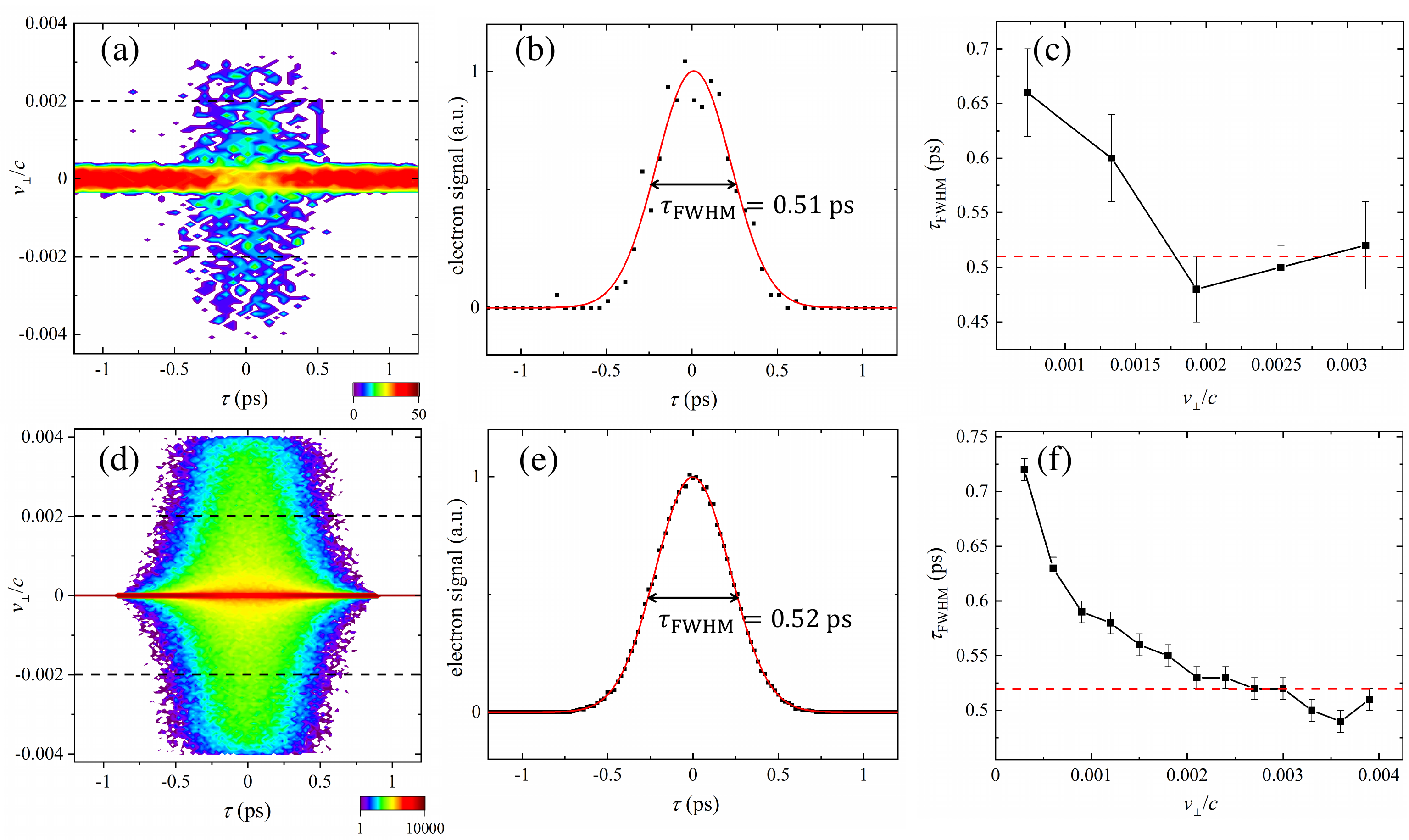}
\caption{\justifying 
Panels (a-c) in the top row show experimental results, while panels (d-f) in the bottom row present the corresponding theoretical calculations for our experimental conditions. (a, d) Transverse velocity distribution of 30~keV electrons as a function of the temporal delay between the electron pulse and the laser pulses forming the intensity grating (logarithmic color scale). The experiment (a) is conducted in a regime without Coulomb interaction with 257.5~nm laser triggering. 
(b, e) Electron pulse profile derived from (a, d) by integrating the electrons with $v_\perp /c$ greater than 0.002 (indicated by dashed lines). The resulting data are fitted with a Gaussian function, and the electron pulse duration is determined from the full-width-at-half-maximum (FWHM) of the fit. (c, f) Electron pulse durations as a function of $v_\perp /c$. The results were obtained by integrating narrow regions of interest along the $v_\perp /c$ axis in (a, d). The error bars indicate standard errors of Gaussian fits. Dashed line represents the electron pulse duration from (b, e).}
\label{fig2}
\end{figure*}

The application of the ponderomotive scattering of electrons via the interaction with optical standing waves to electron pulse duration characterization was first proposed and demonstrated by Hebeisen et al.  \cite{hebeisen2008grating}, where the method is explained in detail. In this approach (see schematics in FIG. \ref{fig1}(b)), electrons traveling through the oscillating electromagnetic field of the optical standing wave are deflected from their original trajectory as they experience a ponderomotive force
\begin{equation}
\textbf{F}\left(\textbf{r}, t\right) = -\frac{e^2 \lambda^2}{8 \pi^2 m_e \epsilon_0 c^3} \nabla I \left(\textbf{r}, t\right).
\end{equation}
\noindent
Here, $e$ is electron charge, $\lambda$ is the wavelenghth of laser, $m_e$ is electron mass, $\epsilon_0$ is vacuum permitivity and $c$ is the speed of light. The ponderomotive force, $\textbf{F}$, is proportional to the gradient of the light intensity of two counterpropagating laser pulses propagating in $x$-direction
\begin{equation}
I(\mathbf{r}, t) = I_0 \cos^2(kx) \cdot g(\mathbf{r}, t),
\end{equation}
where $k$ is wavenumber and $g(\mathbf{r}, t)$ is the laser pulse envelope.

In the original work by Hebeisen et al. \cite{hebeisen2008grating}, the electron pulse duration is determined from the scattered signal by analyzing the time trace. This time trace is defined for each time delay between the electron and laser pulses as the integral of the number of electron counts per pixel weighted by the scattered distance from the beam center. It represents the cross-correlation between the temporal profile of the electron pulse, the temporal shape of the laser pulse, and the transverse spatial profile of the laser pulse as experienced by the electron during its interaction. By deconvolving this signal, the electron pulse duration can be accurately reconstructed. However, the temporal resolution of this method of signal processing is limited both by the duration of the laser pulse and by the time required for the electrons to traverse the laser focus \cite{hebeisen2008grating, gao2012full, chatelain2012ultrafast, baum2013physics}.

\begin{figure*}[t]\centering
\includegraphics[scale=0.4]{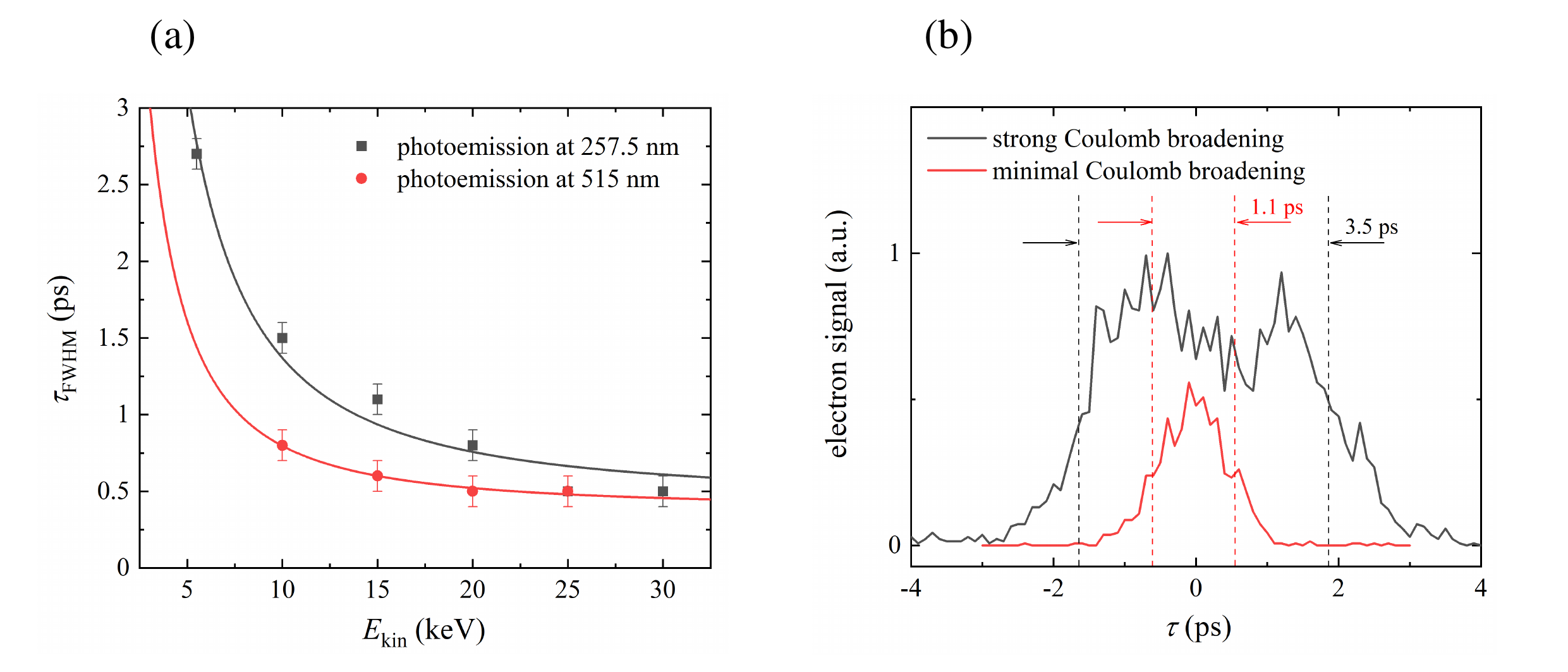}
\caption{\justifying 
(a) Electron pulse durations measured across a broad range of electron energies under experimental conditions designed to minimize the effects of space-charge broadening. The error bars are extended beyond the standard deviation of the Gaussian fit to reflect variations in pulse duration observed across multiple measurements. The black and red curves represent fits to the data using the function from Equation \ref{fit}. (b) The measured electron pulse profile of 15~keV electrons triggered by 257.5~nm laser in the absence (red line) and presence (black line) of space charge effects.
}
\label{fig3}
\end{figure*}

In contrast, we employ a different data processing method, originally introduced in the laser-streaking approach to electron pulse duration measurement \cite{kirchner2014laser}, which has since been widely adopted \cite{plemmons2014characterization, feist2017ultrafast, kozak2018inelastic, kozak2018ultrafast}. FIG. \ref{fig2}(a) shows the measured change in the transverse speed-to-light-speed ratio of 30~keV electrons, $v_\perp /c$, corresponding to the deflection angle of the electrons, as a function of the time delay between the electron pulse and the optical grating. The laser intensity threshold required to achieve higher values of $v_\perp /c$ (i.e., larger scattering angles) increases, indicating that electrons scattered to large angles have to interact with the optical grating within the time window corresponding to high amplitude, effectively shortening the gating time window and thus the response function of the experiment. Ultimately, the response function can approach delta function for filtering of the electrons with only infinitesimaly narrow angular spread close to the maximum scattering angle. However, in our case of electron pulse durations about 500~fs or longer and a laser pulse duration of 220 fs, such strong filtering is unnecessary. The results converge to real electron pulse duration even at small scattering angles, as shown in FIG. \ref{fig2}(c). To determine the electron pulse duration shown in FIG. \ref{fig2}(b), we integrated the number of electrons with $v_\perp /c$ exceeding 0.002 for each time delay. This threshold excludes the broadening of the cross-correlated temporal profile in the near proximity of $v_\perp /c = 0$, which arises from the pedestal in the temporal profile of our real (non-ideal Gaussian) laser pulses. We fitted the resulting electron pulse shape with Gaussian function and determined the electron pulse duration as its full-width-half-maximum (FWHM). FIG. \ref{fig2}(d)-(f) presents numerical simulations corresponding to the experimental results, performed for electrons interacting with pulsed laser beams with Gaussian spatial and temporal envelopes, using a fifth-order Runge-Kutta algorithm to solve the relativistic equation of motion for each electron:
\begin{equation}
\frac{d\left(\gamma m_0{\textbf{v}}\right)}{dt}=q\left({\textbf{E}}+{\textbf{v}}\times{ \textbf{B}}\right).
\end{equation}
\noindent Here $ \textbf{E}$ and $\textbf{B}$ are electric and magnetic fields of two optical pulses with Gaussian envelopes, $m_0$ is electron rest mass,  $q$ is the charge, $\textbf{v}$ is the electron velocity and $\gamma=1/\sqrt{1- \left( v / c \right) ^2}$ is the Lorentz factor.

\section{Results}

We employed the described method to characterize electron pulse durations, $\tau_{\text{FWHM}}$, in an USEM across a~broad range of electron energies, $E_{\text{kin}}$, from 5.5 to 30~keV, as shown in FIG. \ref{fig3}(a). The data were collected under optimal conditions, where the electron emission regime was not limited by Coulomb interactions. Two different photon energies were used for the photoemission pulses to investigate the role of the initial energy spread of the electron spectrum in the final pulse length. For the experimental setup using 257.5~nm (4.8~eV) photoemission pulses, the electron pulse duration at the interaction region ranged from 0.5~ps FWHM for 30~keV electrons to 2.7~ps FWHM for 5.5~keV electrons (results are represented by black squares). To further minimize the electron pulse duration, we changed the laser photoemission wavelength from 257.5~nm (4.8~eV) to 515~nm (2.4~eV). This adjustment brings the photon energy closer to the work function of the Schottky field emitter, reducing the excess energy and, consequently, the initial energy spread $\Delta E_{\text{i}}$\cite{reynolds2023nanosecond}. The results for electron energies between 10 and 25~keV are illustrated in FIG. \ref{fig3}(a), where they are represented as red circles.

The pulse broadening in the absence of space–charge effects is a convolution of the laser pulse duration $\tau_l$ with the temporal electron broadening in the acceleration and field-free drift region caused by initial energy spread $\Delta E_{\text{i}}$ \cite{ gahlmann2008ultrashort}:
\begin{equation}
    \begin{split}
        \tau_{\text{FWHM}} & = \sqrt{\tau_l^2 + \left( \tau_{\text{acc}} + \tau_{\text{drift}} \right)^2} \\
        &\approx \sqrt{ \tau_l^2 + \left( d \sqrt{\frac{m_e}{2 E_i}} \frac{\Delta E_i}{E_{\text{kin}}}  + \frac{l \sqrt{m_e}}{2 \sqrt{2}} \frac{ \Delta E_i }{ E_{\text{kin}}^{3/2}} \right)^2 },
    \end{split}
\end{equation}
\noindent
where $E_i$ is the mean initial energy of electrons and $d$ and $l$ are the lengths of the acceleration region and the field-free region, respectively. Relativistic effects are neglected in this equation for electrons with kinetic energies $E_{\text{kin}}~\leq~30$~keV, and it is assumed that $d~\ll~l$. In the equation it is also assumed that the electric field experienced by the electrons is homogeneous within the acceleration region between the cathode and anode. However, this is not the case for the acceleration system in the USEM, which is more complicated, consists of several electrodes, and does not have a uniform distance $d$ for all final electron kinetic energies. We fit the data using semiempiric equation 
\begin{equation}
\tau_{\text{FWHM}} = \sqrt{ \tau_l^2 + \left( A + \frac{ B }{ E_{\text{kin}}^{3/2}} \right)^2 },
\label{fit}
\end{equation}
\noindent
where we put the acceleration term $A = d \sqrt{\frac{m_e}{2 E_i}} \frac{\Delta E_i}{E_{\text{kin}}}$ as a constant, assuming that $d$ scales linearly with the voltage and, consequently, with the final electron kinetic energy, $d/E_{\text{kin}}=\text{const}$. This assumption is based on the fact that in the electron gun, there are separate electrodes to which the acceleration voltage is applied sequentially. While broadening in the acceleration region is the dominant term for the fast electrons at 30~keV, broadening in the field-free drift region becomes the primary contribution at lower electron energies. Based on the data fit results of parameter $B = \frac{l \sqrt{m_e}}{2 \sqrt{2}}  \Delta E_i $, we can conclude that the initial energy spread $\Delta E_{\text{i}}$ was reduced by a factor of~2 when the laser trigger photon energy was decreased to 2.4~eV. The electron pulse length improved notably, particularly at lower electron energies, where $\Delta E_{\text{i}}$ plays a more substantial role in pulse length broadening due to increased dispersion. Given that we use similar Schottky electron source as Feist et al. \cite{feist2017ultrafast}, we expect $\Delta E_{\text{i}} \approx 0.6 \,\ \text{eV}$ for photoemission at 515~nm. For photoemission at 257.5~nm, our measurements indicate that the initial energy spread is $\Delta E_{\text{i}} \approx 1.2 \,\ \text{eV}$. We note that the fit function is a simplified model and only a rough approximation of the complex pulse broadening mechanism.

FIG. \ref{fig3}(b) compares the measured electron pulse shape for 15~keV electrons triggered by a 257.5~nm laser under two distinct experimental conditions, each characterized by a different number of electrons per pulse. The red curve represents a regime with reduced photoemission laser power, resulting in a lower number of electrons per pulse (approximately 0.1 electrons per pulse detected before the aperture). In this case, Coulomb interactions and space-charge effects are minimized, yielding an optimal temporal duration of 1.1~ps. In contrast, the black curve corresponds to a higher photoemission laser power, producing a greater number of electrons per pulse (approximately 0.8 electrons per pulse detected before the aperture). This leads to increased space-charge effects and a corresponding broadening of the electron pulse duration to about 3.5~ps. It should be noted that the detected electron count differs from the total number of emitted electrons, as some are lost during propagation through the column.

Our experiments showed no significant variation in electron pulse duration across different extractor voltage settings in the range 2700-4000~V. Similarly, the electron beam aperture appeared to have no impact on pulse duration, excluding a significant role of different electron trajectories within the beam profile.

\begin{figure}[t]\centering
\includegraphics[scale=0.4]{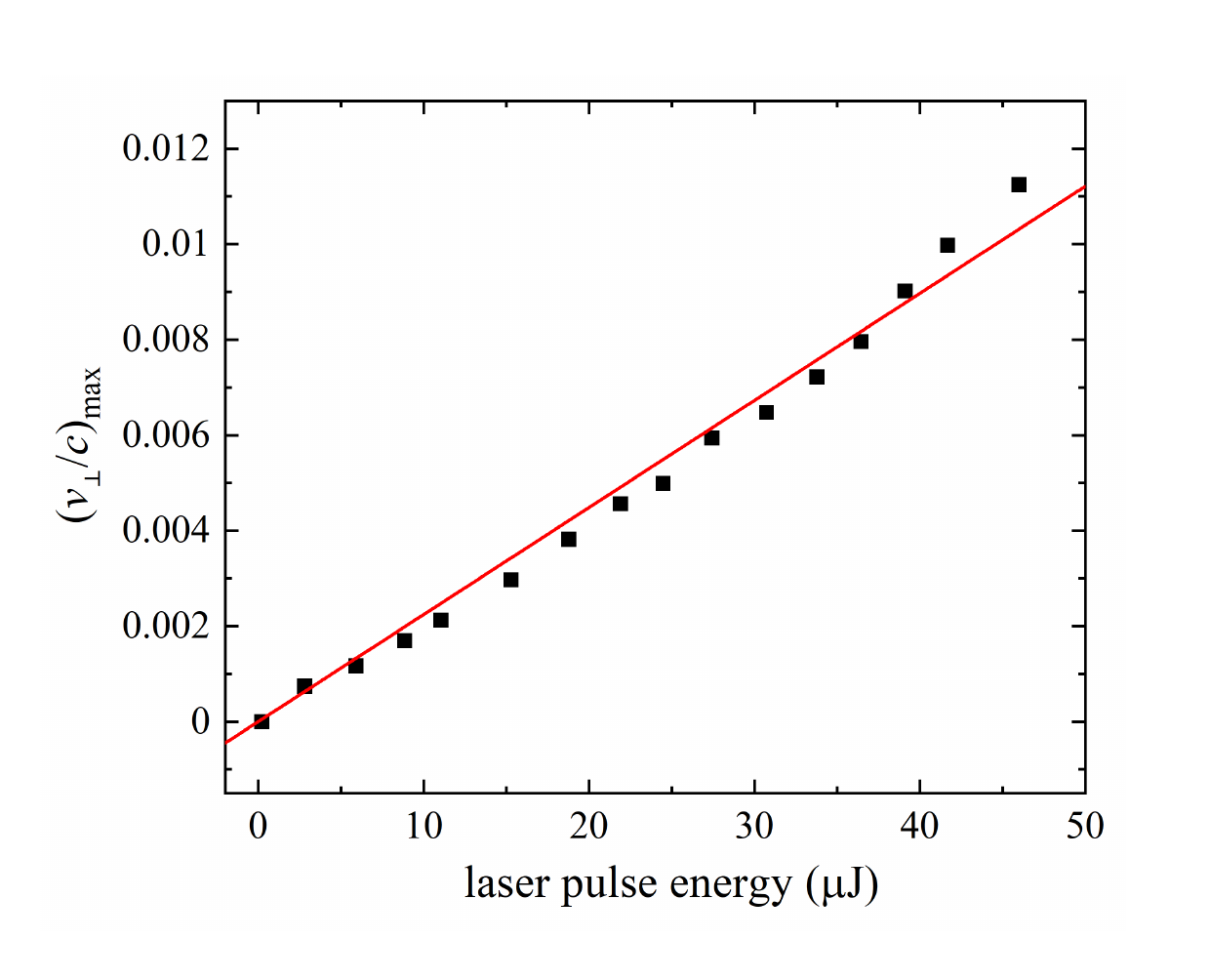}
\caption{\justifying 
The transverse speed-to-light-speed ratio $ {\left( v_\perp /c \right)}_{\text{max}} $, corresponding to the maximum electron deflection, is shown as a function of the laser pulse energy used to create the optical grating. The measurements were performed for 30~keV electrons.
}
\label{fig4}
\end{figure}

We verified that the maximal electron scattering angle (and therefore the corresponding transverse velocity ratio $ {\left( v_\perp /c \right)}_{\text{max}} $) scales linearly with the intensity of the optical grating, showing no indications of saturation, as illustrated in FIG. \ref{fig4} (for laser pulse energy). This confirms that the displacement experienced by electrons as a result of the ponderomotive force during the interaction is negligible compared to that experienced during one period of the optical grating. Consequently, the force acting on an electron can be approximated as acting along a straight line, and the laser pulse energy used in our measurements does not have any influence on the measured electron pulse duration \cite{hebeisen2008grating}. If the linear condition were not satisfied, the resulting temporal durations may appear artificially longer.

The peak light intensity required to induce the angular deflection of electrons must be sufficiently high to overcome the angular distance between two neighboring pixels on the detector, ensuring that the scattering signal can be observable. This, along with the spot size of the electron beam on the detector, establishes a limit on the minimum pulse energy needed to implement the method. Based on the data in FIG. \ref{fig4}, the minimum laser energy required to observe scattering in our setup, with a 170~mm distance between the interaction region and the detector, is approximately 500nJ. For most of our measurements, a laser pulse energy of 20~$\mu$J was used to generate the optical grating.

\section{Conclusions}

In the past, advancements in the direct measurement of ultrashort electron pulse durations in UEMs have predominantly focused on setups equipped with high-resolution spectrometers, where the inelastic interaction between electrons and light has been utilized to characterize the electron pulses. In USEMs, where such spectrometers are not yet broadly implemented, the electron pulse durations have been typically estimated based on theoretical calculations.

Here, we demonstrate the implementation of an all-optical method for the direct temporal characterization of femtosecond electron pulses within an USEM. This method is based on the electron scattering on an optical standing wave and can operate over a broad range of electron energies (1–30~keV). Furthermore, we show that the electron pulse duration can be significantly reduced by using a photoemission wavelength of 515nm, compared to the more commonly used 257.5nm.

\section*{Data availability statement}

All data supporting the findings of this study are openly available in the Zenodo repository at \href{https://doi.org/10.5281/zenodo.15573403}{https://doi.org/10.5281/zenodo.15573403}.

\begin{acknowledgments}
The authors acknowledge funding from the Czech Science Foundation (project 22-13001K), Charles University (SVV-2024-260720, PRIMUS/19/SCI/05, GAUK 216222) and the European Union (ERC, eWaveShaper, 101039339). Views and opinions expressed are however those of the author(s) only and do not necessarily reflect those of the European Union or the European Research Council Executive Agency. Neither the European Union nor the granting authority can be held responsible for them. This work was supported by TERAFIT project No. CZ.02.01.01/00/22\_008/0004594 funded by OP JAK, call Excellent Research.
\end{acknowledgments}

\bibliography{biblio}

\end{document}